\documentclass[preprint2]{aastex}

\def\p{$\pm$}

\def\ab{$\sim$}
\def\kms{km s${}^{-1}$}
\def\beam{beam$^{-1}$}

\def\kms{km s${}^{-1}$}

\def\a{$\alpha$}
\def\ab{$\sim$}

\def\p{$\pm$}

\def\ad{$\alpha$,$\delta$$_{(J2000)}$}
\def\beam{beam$^{-1}$}
\def\x{$\times$}
\def\T{T$_e^*$}

\def\mdot{\.{M}}
\def\msun{M$_{\sun}$}
\def\yr{yr$^{-1}$}
\def\vinf{v$_\infty$}

\begin{document}

\title{VLA Detection of the Ionized Stellar Winds Arising from Massive Stars in the Galactic Center Arches Cluster}
\author{Cornelia C. Lang\altaffilmark{1,2}, W. M. Goss\altaffilmark{1}, Luis F. Rodr\'{\i}guez\altaffilmark{3}}
\altaffiltext{1}{National Radio Astronomy Observatory, Box 0, Socorro,
NM 87801; CCL's current address: Astronomy Program, LGRT B-517O, University of Massachusetts, Amherst, MA 01003, email: clang@ocotillo.astro.umass.edu}
\altaffiltext{2}{Division of Astronomy, 8371 Math Sciences Building, Box
951562, University of California at Los Angeles, LA, CA 90095-1562} 
\altaffiltext{3}{Instituto de Astronom\'{\i}a, UNAM, Campus Morelia, Apdo. Postal 3-72, Morelia, Michoac\'{a}n 58089, M\'{e}xico}

\begin{abstract}
The Galactic center Arches stellar cluster, detected and studied until now only in the near-infrared, is comprised of at least one hundred massive (M$_\star \geq$ 20 M$_\odot$) stars. Here we report the detection at centimeter wavelengths of radio continuum emission from eight radio sources associated with the cluster. Seven of these radio sources have rising spectral indices between 4.9 and 8.5 GHz and coincide spatially with the brightest stars in the cluster, as determined from JHK photometry and Br$\alpha$ and Br$\gamma$ spectroscopy. Our results confirm the presence of powerful ionized winds in these stars. The eighth radio source has a nonthermal spectrum and its nature is yet unclear, but it could be associated with a lower mass young star in the cluster.
\end{abstract}


To appear in The Astrophysical Journal Letters

\section{Introduction}

High-resolution observations of the Galactic center at near-infrared wavelengths first revealed the remarkable Arches stellar cluster located at $\ell$=0\fdg12, b=0\fdg02, \ab11\arcmin~(or 27.5 pc) in projection from Sgr A$^*$ (Nagata et al. 1995; Cotera et al. 1996). Here, we assume a distance to the Galactic center of 8.0 kpc (Reid 1993). Nagata et al. (1995) identified the 14 brightest stars in this cluster with JHK photometry and Br$\gamma$ and Br$\alpha$ hydrogen recombination lines, showing that these stars had the characteristic colors and emission lines of Wolf-Rayet (WR) and He I emission-line stars. The strength of the Br$\alpha$ and Br$\gamma$ emission lines suggests that these stars are losing mass at rates of \mdot\ab2$\times$10$^{-5}$ \msun~\yr. Using K-band spectroscopy, Cotera et al. (1996) also found 12 stars in the cluster with spectra consistent with late-type WN/Of types, and which have \mdot\ab1$-$20 $\times$ 10$^{-5}$ M$_{\sun}$ \yr, and \vinf\ab 800$-$1200 \kms.

Recent observations of the Arches cluster indicate that the number of massive stars is much greater than suggested by the initial studies. Serabyn et al. (1998) estimate that at least 100 cluster members have masses $>$ 20 M$_{\sun}$, and Figer et al. (1999) calculate that the total cluster mass exceeds 10$^4$ M$_{\sun}$ with as many as 160 O-stars and a stellar density of $\rho$ \ab3$\times$10$^5$ M$_{\sun}$ pc$^{-3}$. These parameters make the Arches cluster one of the densest and most massive young stellar clusters in the Galaxy. In addition, the ionizing flux (estimated to be a few \x~10$^{51}$ s$^{-1}$) is sufficient to ionize the surrounding Arched Filaments H II complex (Serabyn et al. 1998; Lang et al. 2001). The relative placement of the Arches cluster and the filamentary nebular features is depicted in Figure 1.

Thermal radio emission from ionized stellar winds is detectable at radio wavelengths and arises from the outer parts of the wind envelope. The classic theory of Panagia \& Felli (1975) and Wright \& Barlow (1975) predicts that the spectrum of radio emission from a stellar wind is $\propto$ $\nu^{+0.6}$ for a spherically symmetric, isothermal, stationary wind expanding at a constant velocity.  Previous surveys made with the Very Large Array (VLA) of the National Radio Astronomy Observatory\footnotemark\footnotetext{The National Radio Astronomy Observatory is a facility of the National Science Foundation operated under cooperative agreement by Associated Universities, Inc.} have detected radio emission from OB supergiant and Wolf-Rayet stellar winds (Abbott et al. 1986; Bieging et al. 1989). Recently, Lang et al. (1999) detected six radio point sources in the vicinity of a similar Galactic center cluster, the Quintuplet, which have spectra consistent with the theoretical predictions and close positional correspondances with known near-infrared stellar sources. Here, we present multi-frequency VLA observations of the Arches cluster in an attempt to detect stellar wind emission from the individual stars known to be members of this dense cluster. 

\section{VLA Continuum Observations}
VLA multi-frequency continuum observations were made of the Arches cluster in July 1999 (A-array) at 4.9 GHz and 8.5 GHz (4 hours each), and in October 1999 (BnA array) at 43.3 GHz (3 hours). The phase center of all observations was the position given for the bright, central source in Nagata et al. (1995): \ad=17 45 50.41, $-$28 49 21.8.  Standard procedures for calibration, editing, and imaging were carried out using the {\itshape AIPS} software of NRAO. At all wavelengths, 3C286 was used for flux density calibration, and at 8.5 and 4.8 GHz, 1751-253 was used as a phase calibrator. At 43 GHz, fast switching between the bright source Sgr A$^*$ and the phase center of the observations (an angular distance of 12\farcm5) was carried out on a 90 second cycle in order to closely track the phase variations. The resulting spatial resolutions and rms noise levels in the images are 0\farcs79 \x~0\farcs33, PA=12\fdg8 and 45 $\mu$Jy \beam~at 4.9 GHz; 0\farcs41 \x~0\farcs16, PA=2\fdg8 and 25 $\mu$Jy \beam~at 8.5 GHz; and 0\farcs29 \x~0\farcs07, PA=49\fdg2 and 300 $\mu$Jy \beam~at 43.3 GHz.

\begin{figure}[t!]
\plotone{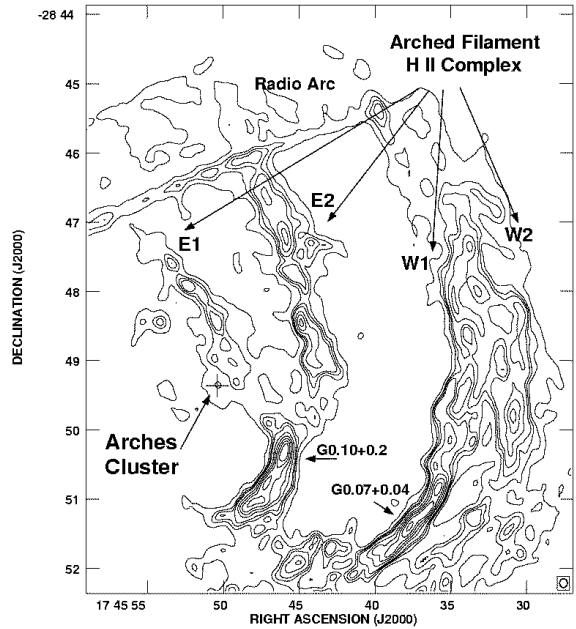}
\caption{Location of Arches stellar cluster and the surrounding Arched Filament H II nebula. The contours represent VLA 8.5 GHz continuum emission at levels of 7.5, 15, 20, 25, 37.5, 50, 75, 87.5 mJy \beam, with a resolution of 7\farcs8 \x~6\farcs6, PA=$-$0\fdg7, from Lang et al. (2001).}
\end{figure}                                                                                                                  

\section{Results}
Figure 2 shows the 8.5 GHz image of the radio sources within a 16\arcsec~(0.7 pc) region centered on the bright source at the cluster center. The crosses represent the positions of 9 of the 14 stars in Nagata et al. (1995) which showed Br$\gamma$ and Br$\alpha$ recombination lines. These positions have been shifted by 0\fs05 (0\farcs66) to correct for what appears to be a systematic offset to the West. At 8.5 GHz, eight sources are detected above a 6$\sigma$ level and are referred to as AR1$-$8. Seven of the sources (with the exception of AR6) appear to be associated with the stellar positions plotted in Figure 2. AR7 also shows a larger displacement ($\Delta$$\alpha$=0\farcs5) from the corresponding stellar position than the rest of the radio sources. However, the absolute uncertainty in the near-infrared positions is \ab2\arcsec~or less, so an association is quite possible within these limits. Since the a priori probability of finding a 8.5 GHz background source with flux density greater than 0.1 mJy in the region considered is \ab0.005, we conclude that all the radio sources detected are indeed associated with the Galactic center Arches cluster.
 
\begin{figure}[t!]
\plotone{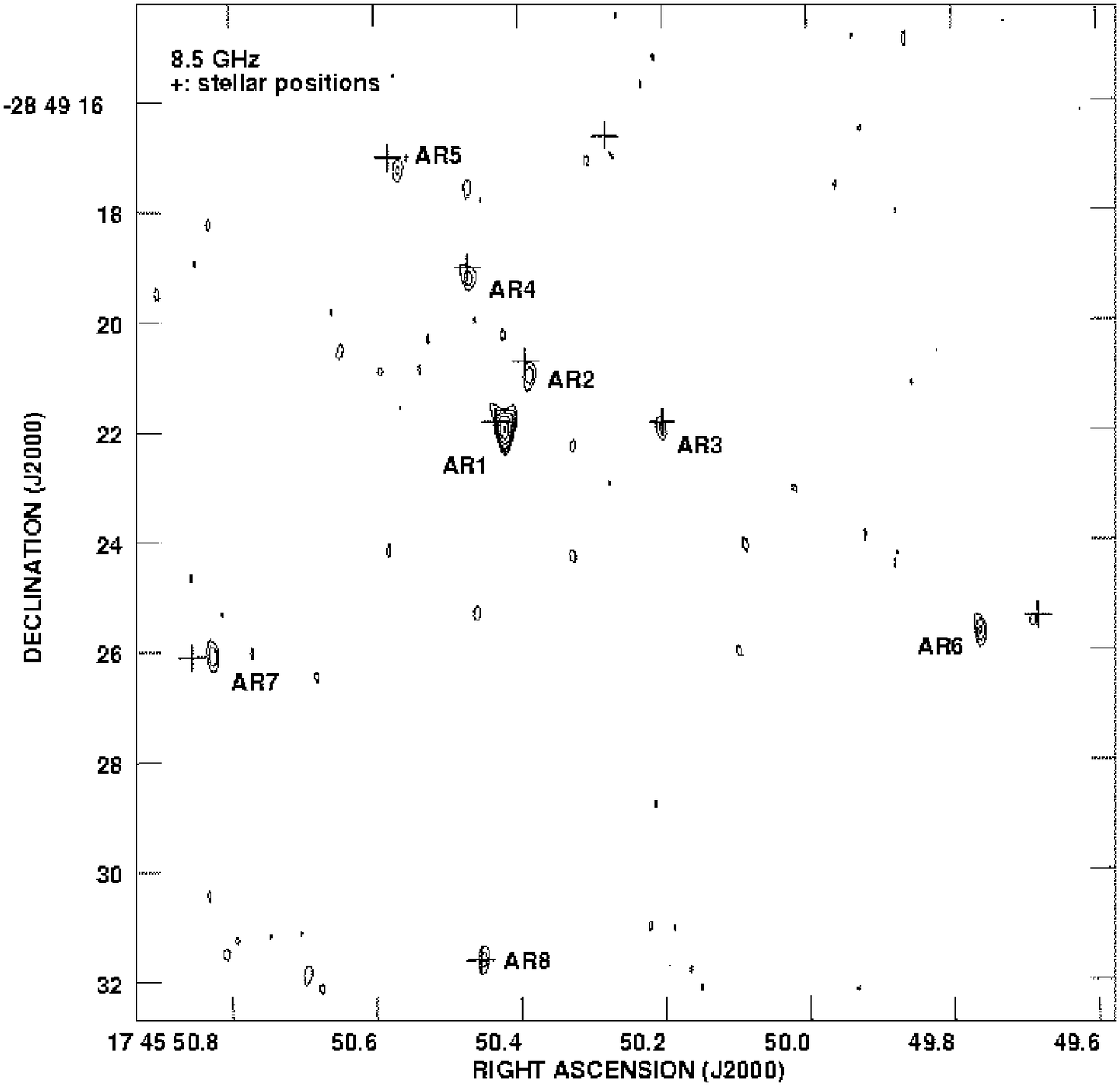}
\caption{VLA 8.5 GHz continuum image of the inner 16\arcsec~(0.7 pc) of the Arches cluster, with a resolution of 0\farcs41 \x~0\farcs16, PA=2\fdg8. This image has been primary beam corrected and has an rms of 25 $\mu$Jy \beam. The contour levels of 0.075, 0.15, 0.25, 0.5, 1.0, 1.5, 1.75 mJy \beam correspond to 3, 6, 10, 20, 40, 60, 70 $\sigma$ and the crosses correspond to the positions of the nine near-infrared sources of Nagata et al. (1995) detected in Br$\gamma$ and Br$\alpha$ lines. These positions are only known to an absolute accuracy of \ab2\arcsec~(or 0\fs15), and here have been shifted by 0\farcs66 (or 0\fs05) to the West to align the sources more closely.}
\end{figure}

The positions and properties of AR1$-$8 are listed in Table 1. Four of these sources are detected in the 4.9 GHz observations (AR1, AR4, AR6, AR7), while the central source AR1 is additionally detected in the 43.3 GHz observations. At 43.3 GHz, AR1 has a flux density of 3.1\p0.1 mJy, consistent with the 4.9/8.5 GHz spectral index ($\alpha$\ab+0.35). Spectral indices between 4.9/8.5 GHz are calculated for the sources detected at 4.9 GHz. Upper limits for the flux densities at 4.9 GHz can be estimated for the remaining sources, with resulting lower limits for the spectral indices. Seven of the sources show rising spectra in the range of +0.3 to +0.9 with the exception of AR6, which has a nonthermal spectral index (\a=$-$0.7). The sources AR2$-$8 are point sources with angular sizes $\leq$ 0\farcs25. The deconvolved size of AR1 at 8.5 GHz is 0\farcs11 (0.005 pc), implying a brightness temperature of \ab3000 K, a reasonable value for an ionized wind. 

\section{Discussion}
\subsection{The Nature of the Radio Sources}
The spectral indices of the detected sources are consistent with values expected from stellar wind emission. Several of the sources have spectra which are flatter ($\alpha$~\ab+0.3) than the predicted $\alpha$=+0.6, and AR6 has a negative spectral index ($\alpha$~\ab$-$0.7). However, these spectra do not rule out stellar wind detections. About 25\% of stellar winds arising from massive stars have been found to have flat or nonthermal spectra in the radio regime (Bieging, Abbott \& Churchwell 1989). The precise mechanism that produces these nonthermal spectra is not established, but internal shocks in the wind of a single star (White 1985) or the presence of interacting winds in close binary systems have been proposed (Contreras \& Rodr\'{\i}guez 1999). For a typical late-type WN star at the Galactic center, the expected flux density at 8.5 GHz arising from a stellar wind is \ab250 $\mu$Jy based on the assumed values of \vinf\ab1000 \kms, \mdot\ab5$\times$10$^{-5}$ \msun~\yr~and T\ab10$^4$ K. This value is similar to the flux densities we measure at 8.5 GHz for AR2$-$8. 
Figure 2 also shows an excellent positional correspondence between the radio sources and the stellar sources detected by Nagata et al. (1995). Table 2 lists the spectral types of these stars as determined by Cotera et al. (1996). The stars have been classified as late-type WN and Of stars, all of which are expected to be losing mass at high rates from their surfaces. As noted above, such stars often exhibit flattened or nonthermal spectral indices due to the presence of a nonthermal component. Therefore, the range of spectra we detect is consistent with predictions for late-type WN stars. 

The radio source AR6 is displaced by $\Delta$$\alpha$=1\arcsec~from the corresponding stellar position of Nagata et al. (1995). At 8.5 GHz, there is a 75 $\mu$Jy source (3$\sigma$) closer to the stellar position, which may be the correct identifcation. The stellar source at this position has been classified as a WN8 star by Cotera et al. (1996). If this displacement is real, we have then to consider the possibility that AR6 is associated with another star in the cluster. T Tauri stars are sometimes found to have associated nonthermal emission, usually interpreted as gyrosynchrotron produced by electrons accelerated in situ by magnetic reconnection flares near the stellar surface (Andr\'e et al. 1988; Rodr\'{\i}guez, Anglada, and Curiel 1999). If this is the case, we may be detecting for the first time a young, low mass member of the cluster. 

\subsection{Mass Loss Rates}
The flux densities at 8.5 GHz were used to calculate a mass-loss rate based on Panagia \& Felli (1975):

\begin{small}
\begin{equation}
\frac{\dot{M}}{10^{-5} M_\odot~yr^{-1}} = 0.52 \left(\frac{S_{8.5}}{mJy}\right)^{3/4} \left(\frac{v_{\infty}}{10^3~km~s^{-1}}\right) \left(\frac{d}{kpc}\right)^{3/2} 
\end{equation}
\end{small}
where S$_{8.5}$ is the flux density of the source at 8.5 GHz, \vinf~is the terminal velocity of the wind, and d is the distance to the source (8 kpc). We have assumed an electron temperature of \T=10$^4$ K, Z=1, and a mean molecular weight, $\mu$=2, due to the enrichment in heavy elements of the late-type WN stars (Leitherer et al. 1997). The mass loss rate of AR1 has the largest value, \mdot=1.7 \x~10$^{-4}$ \msun~yr$^{-1}$, comparable only to stars with extreme values for \mdot, such as in the R136 stellar cluster at the center of 30 Doradus (de Koter et al. 1997). The values of \mdot~for AR2$-$8 are in the range of 3$-$4.5 \x~10$^{-5}$ \msun~\yr, consistent with values found for for late type WN stars by Leitherer et al. (1997): \mdot\ab2.5$-$3.9 $\times$10$^{-5}$ \msun~yr$^{-1}$. 

\subsection{The Arches Cluster and the Interstellar Medium}

The rich population of massive stars in the Radio Arc region, which includes both the Arches and Quintuplet clusters (Figer et al. 1999), suggests that the interstellar environment should be strongly influenced by the energetic processes related to such massive star formation and evolution. Ionization of the molecular material in this region and formation of the unusually shaped Arched Filaments and Sickle H II regions can be accounted for by the ionizing fluxes of each cluster (Lang et al. 1997, 2001). In addition, interactions between the winds in the densely-packed Arches cluster may give rise to a shock-heated diffuse ``cluster wind'', which could have a temperature as high as 10$^7$ K (Cant\'o, Raga \& Rodr\'{\i}guez 2000). This emission is predicted to be detectable in the X-ray regime with the {\it Chandra X-ray Observatory}. Studies of the interaction between the massive stars in the Arches and the surrounding interstellar environoment are currently underway at a number of complementary wavelengths. 

\section{Conclusions} 

We present the following results from our 4.9, 8.5, and 43.3 GHz observations of the Arches Cluster:

(1) Eight radio sources are detected at 8.5 GHz, four sources at 4.9 GHz, and one of them is detected also at 43.3 GHz. Seven of the eight detections show rising spectral indices between 4.9 and 8.5 GHz in the range +0.3 to +0.6, while one source shows a nonthermal spectrum ($-$0.7). All of the detections are consistent with stellar wind emission from mass-losing stars. However, the nonthermal source may be tracing a young, low mass member of the cluster. This is the first detection of emission associated with the members of the Arches cluster at a band other than the near-infrared.

(2) Seven of the eight radio sources are coincident in position with the Br$\gamma$ and Br$\alpha$ emission-line stellar sources of Nagata et al. (1995), which correspond to the late WN and Of stellar types as classified by Cotera et al. (1996). These positional correspondences further confirm that the ionized winds from stars in the Arches cluster are detected.

(3) Based on the radio flux density, AR1 has a mass-loss rate of \ab2$\times$10$^{-4}$~\msun~yr$^{-1}$,  comparable only to that of the stars in 30 Doradus (de Koter et al. 1997). Mass-loss rates for AR2$-$8 are found to be in the range 3$-$4.5$\times$10$^{-5}$ \msun~\yr, similar to the large mass loss rates for other late-type WN systems (Leitherer et al. 1997).

\clearpage
\begin{deluxetable}{lcccccc}
\tablecaption{Parameters of the Radio Sources (AR1-8) in the Arches Cluster} 
\tablewidth{0pt}
\tablehead{\colhead{Source} &
\colhead{RA (J2000)\tablenotemark{\dagger}} &
\colhead{DEC (J2000)\tablenotemark{\ddagger}} &
\multicolumn{2}{c}{Flux Density (mJy)} &
\colhead{Spectral}&
\colhead{\mdot}\\
\cline{4-5}
\colhead{Name}&
\colhead{(h m s)}&
\colhead{(\arcdeg~\arcmin~\arcsec)}&
\colhead{$\nu$=8.5 GHz} & 
\colhead{$\nu$=4.9 GHz} &
\colhead{Index}&
\colhead{(\msun~\yr)}}
\tablecolumns{7}
\startdata
AR1&17 45 50.42&$-$28 49 22.0&1.70\p0.05&1.40\p0.03&+0.35\p0.04&1.7\x10$^{-4}$\\
AR2&17 45 50.39&$-$28 49 21.0&0.23\p0.02&$<$0.13&$>$+0.9&3.9\x10$^{-5}$\\
AR3&17 45 50.20&$-$28 49 22.0&0.17\p0.02&$<$0.13&$>$+0.4&3.2\x10$^{-5}$\\
AR4&17 45 50.47&$-$28 49 19.2&0.23\p0.02&0.16\p0.03&+0.65\p0.13&3.9\x10$^{-5}$\\
AR5&17 45 50.57&$-$28 49 17.2&0.16\p0.02&$<$0.13&$>$0.3&3.0\x10$^{-5}$\\
AR6&17 45 49.76&$-$28 49 25.7&0.27\p0.02&0.40\p0.03&$-$0.71\p0.08&4.5\x10$^{-5}$\\
AR7&17 45 50.83&$-$28 49 26.1&0.25\p0.02&0.21\p0.03&+0.31\p0.05&4.2\x10$^{-5}$\\
AR8&17 45 50.45&$-$28 49 31.7&0.20\p0.02&$<$0.13&$>$+0.7&3.6\x10$^{-5}$\\
\enddata
\tablenotetext{\dagger}{The positional accuracy in RA is \p0\fs01}
\tablenotetext{\ddagger}{The positional accuracy in DEC is \p0\farcs1}
\end{deluxetable}
\begin{deluxetable}{ccccc}
\tablewidth{0pt}
\tablecaption{Near-Infrared Counterparts of Arches Radio Sources} 
\tablehead{\colhead{Radio} &
\colhead{Nagata et al.}&
\colhead{Cotera et al.}&
\colhead{Spectral}\\
\colhead{Source}&
\colhead{Source \#}&
\colhead{Source \#}&
\colhead{Type\tablenotemark{*}}}
\tablecolumns{5}
\startdata
AR1&8&8&Of/WN9\\
AR2&7&6&WN8/9\\
AR3&4&9&Of/WN9\\
AR4&10&5&Of/WN9\\
AR5&11&2&WN8\\
AR6&1\tablenotemark{\dagger}&13&WN8\\
AR7&14&11&Of/WN9\\
AR8&9&$-$&$-$\\
\enddata
\tablenotetext{*}{From Cotera et al. (1996).}
\tablenotetext{\dagger}{Identification with near-infrared source not firm.}
\end{deluxetable}

\end{document}